\documentclass[journal]{IEEEtran}

\usepackage{cite}

\ifCLASSINFOpdf
\usepackage[pdftex]{graphicx}
\else
\fi
\usepackage{amsmath}
\ifCLASSOPTIONcompsoc
\usepackage[caption=false,font=normalsize,labelfont=sf,textfont=sf]{subfig}
\else
\usepackage[caption=false,font=footnotesize]{subfig}
\fi
\usepackage{silence}
\usepackage{caption}														
\usepackage{booktabs}																				
\usepackage[dvipsnames]{xcolor}															

\usepackage{pgfplots,pgf,tikz,xparse,tikz-3dplot}						
\usepackage{tikz-timing}[2009/12/09]
\usepackage[tight]{units}				                            
\pgfplotsset{compat=1.9}

\usepackage{pgf}
\usepackage{tikz}
\usepackage{xparse}
\usetikzlibrary{calc, through, matrix, shadings, arrows, backgrounds, arrows.meta, shapes, automata, shadows}
\usetikzlibrary{decorations, decorations.markings, decorations.pathreplacing, decorations.pathmorphing}
\usetikzlibrary{pgfplots.units, spy, patterns, positioning, intersections, fadings, mindmap, trees}
\usetikzlibrary{decorations, decorations.pathreplacing, decorations.pathmorphing}

\pgfplotsset{footnotesize, major grid style={thick}, y tick label style={/pgf/number format/1000 sep=}, use units, grid=both,
	every axis legend/.append style={font=\footnotesize, rounded corners=2pt, legend cell align=left}}
\pgfplotsset{tick label style={font=\footnotesize}, label style={font=\footnotesize}, legend style={font=\footnotesize}}
\pgfplotsset{every axis/.append style={line width=.5pt}}

\pgfdeclarelayer{background}
\pgfsetlayers{background,main}

\tikzstyle{Linie}  = [thick, double distance=.5mm] 
\tikzstyle{Pfeil1} = [thick, decoration={markings, mark=at position 1 with {\arrow[semithick, fill=white]{triangle 60}}},
double distance=.5mm, shorten >= 2mm, preaction = {decorate},
postaction = {draw, line width=.5mm, white, shorten >= 1mm, shorten <= 0mm}]

\tikzstyle{branch}=[draw, fill, shape=circle, minimum size=3pt, inner sep=0pt]

\definecolor{LightBlue1}     {HTML}{BFEFFF}

\colorlet{NewOrange}{orange!70!black}

\newcommand{\mm}{\text{-}}
\newcommand{\sm}[1]{\scalebox{0.7}{#1}}
\newcommand{\sn}[1]{\scalebox{0.95}{#1}}

\newcommand{\ha}{\hspace{-1mm}}

\newcommand{\hd}{\hspace{-4mm}}

\begin{document}

\title{Adaptive Extended Kalman Filter (ROSE-Filter) \\ for Positioning System}


\author{Reiner~Marchthaler (orcid: 0000-0001-9108-3072)
\thanks{R.~Marchthaler was with the Department of Information Technology, University of Applied Sciences, Esslingen,
Flandernstr. 101, 73732 Esslingen, Germany, e-mail see: https://www.hs-esslingen.de/en/staff/reiner-marchthaler/.
}
}

%
%

\markboth{arXiv.org, open-access archive, computer science, August~2021}%
{
}
%


\maketitle 																														

\begin{abstract}
\IEEEPARstart{T}{his} paper illustrates the way for estimating position and orientation of a vehicle with an Extended Kalman Filter (EKF).
For this purpose a non-linear model is designed and an adaptive calculation of measurement noise covariance matrix is used, a so called ROSE-Filter (\underline{R}apid \underline{O}ngoing \underline{S}tochastic covariance \underline{E}stimation-Filter).
Input of the system is the measured position from a two dimensional position system.
Estimated is the pose (position and orientation) and the velocity of the vehicle.
\end{abstract}


%
\IEEEpeerreviewmaketitle

\section{Introduction}

The exact pose (position and orientation) of objects is required for a lot of applications. 
This pose is base of controlling vehicles and creating a model of the environment.  

Sensors, like GPS just frequently measured the position not accurately.
Therefore it is necessary to filter the position and estimate the orientation of the object.

The Kalman filter is a popular approach to solve this topic.
This filter requires to first define a model to estimate the pose.
There are different applications which are based on linear models and non-linear models.
Each approach has its own advantages and disadvantages. Linear models have problems estimating the orientation of an object.
However non-linear models can estimate the pose but are less stable as linear models. Hence a stable non-linear model is required. 

This paper focuses on vehicles, like cars or robots, based on a planar vehicle model.
The planar vehicle model uses the steering for changing the direction of the vehicle to describe the x- and y- position of the center of gravity and the orientation of the vehicle.
Such kind of modeling is described in more detail in \cite{Lugner.2019} and \cite{Schramm.2014}.

\section{Modeling}
\subsection{Nonlinear model}
\label{ref:NonlinearModel}

The trajectory of objects based on a planar vehicle model can be described within a small time step as circular trajectory with radius $R$ around the center of the circle $(x_c,y_c)$. 

The position of the object at time $t=0$ is given by
\begin{IEEEeqnarray}{lCl}
	x\sm{(0)} &=& x_c + R \!\cdot\! \cos(\varphi\sm{(0)}) \label{eq:x0} \\
	y\sm{(0)} &=& y_c + R \!\cdot\! \sin(\varphi\sm{(0)}) \label{eq:y0}
\end{IEEEeqnarray}

and at time $t=1$ given by
\begin{IEEEeqnarray}{lCl}
	x\sm{(1)} &=& x_c + R \!\cdot\! \cos(\varphi\sm{(1)}) \label{eq:x1} \\
	y\sm{(1)} &=& y_c + R \!\cdot\! \sin(\varphi\sm{(1)}) \label{eq:y1}
\end{IEEEeqnarray}

Solve equations \eqref{eq:x0} for $x_c$ and \eqref{eq:y0} for $y_c$ and substitute into \eqref{eq:x1} and \eqref{eq:y1}. Thus,
\begin{IEEEeqnarray}{lCl}
	x\sm{(1)} &=& R \!\cdot\! \cos(\varphi\sm{(1)}) + x\sm{(0)} - R \!\cdot\! \cos(\varphi\sm{(0)}) \nonumber \\
	&=& x\sm{(0)} + R \!\cdot\! \bigl(\cos(\varphi\sm{(1)}) - \cos(\varphi\sm{(0)})\bigr) \nonumber \\
	&=& x\sm{(0)} - 2 \!\cdot\! R \!\cdot\! \sin\bigl(\tfrac{\varphi\sm{(1)} + \varphi\sm{(0)}}{2} \bigr) \!\cdot\! \sin\bigl(\tfrac{\varphi\sm{(1)} - \varphi\sm{(0)}}{2} \bigr) \label{eq:x1b} \\
	y\sm{(1)} &=& R \!\cdot\! \sin(\varphi\sm{(1)}) + y\sm{(0)} - R \!\cdot\! \sin(\varphi\sm{(0)}) \nonumber \\
	&=& y\sm{(0)} + R \!\cdot\! \bigl(\sin(\varphi\sm{(1)}) - \sin(\varphi\sm{(0)})\bigr) \nonumber \\
	&=& y\sm{(0)} + 2 \!\cdot\! R \!\cdot\! \cos\bigl(\tfrac{\varphi\sm{(1)} + \varphi\sm{(0)}}{2} \bigr) \!\cdot\! \sin\bigl(\tfrac{\varphi\sm{(1)} - \varphi\sm{(0)}}{2} \bigr) \label{eq:y1b} 
\end{IEEEeqnarray}

\vspace*{5mm}
\begin{figure}[!t]	
	\begin{center}
		\begin{tikzpicture}
			\def\R{4}
			\def\PhiA{15}
			\def\PhiB{80}
			\def\PhiC{0.5*\PhiB+0.5*\PhiA}
			
			\coordinate (Ursprung) at (1.5,0.7);
			\coordinate (A) at ($(Ursprung)+(\PhiA:\R)$);
			\coordinate (B) at ($(Ursprung)+(\PhiB:\R)$);
			\coordinate (C) at ($(Ursprung)+(\PhiC:\R)$);
			
			\draw[->,>={Stealth[length=3mm]},thick] (-.3,0) -- (7,0) node[below left,yshift=-1mm]{x};
			\draw[->,>={Stealth[length=3mm]},thick] (0,-.3) -- (0,6) node[below left]{y};
			
			\draw[rotate=\PhiA,fill=lightgray,draw=darkgray,rounded corners=1.5mm] ($(A)-(.3,.5)$) rectangle ($(A)+(.3,.5)$);
			\draw[rotate=\PhiB,fill=lightgray,draw=darkgray,rounded corners=1.5mm] ($(B)-(.3,.5)$) rectangle ($(B)+(.3,.5)$);
			
			\draw[darkgray,dashed] (Ursprung)+(-15:\R) arc(-15:117:\R);
			\draw[darkgray,very thick,->,>={Stealth[length=2.5mm]}] (A) arc(\PhiA:\PhiB:\R);
			
			\draw[darkgray,dashed] (Ursprung) --++ (5,0);
			
			\draw[red,dashed] (Ursprung) -- node[above right,shift={(2mm,1mm)}]{$R$}(A);
			\draw[fill=red] (Ursprung) --++ (0:0.4*\R) arc(0:\PhiA:0.4*\R);
			\fill[red] (Ursprung)+(0.5*\PhiA:0.4*\R) node[right]{$\varphi\sm{(0)}$};
			\fill[red]  (A) circle(1.2mm) node[right, xshift=2.5mm]{$(x\sm{(0)},y\sm{(0)})$};
			
			
			\draw[blue,dashed] (Ursprung) -- node[left]{$R$} (B);
			\draw[fill=blue,semitransparent] (Ursprung) --++ (0:0.2*\R) arc(0:\PhiB:0.2*\R);
			\fill[blue] (Ursprung)+(0.7*\PhiB:0.2*\R) node[right,xshift=1mm]{$\varphi\sm{(1)}$};
			\fill[blue] (B) circle(1.2mm) node[above right,xshift=1mm,yshift=1.5mm]{$(x\sm{(1)},y\sm{(1)})$};
			
			\fill[black](Ursprung) circle(1.2mm) node[left, xshift=-1mm]{$(x_c,y_c)$};
			\fill[black](C) node[above right]{$\sm{$\Delta$} s\sm{(0)}$};
			
		\end{tikzpicture}
		\caption{Modeling trajectory based on circular path}
		\label{fig_model_circularpath} 
	\end{center}
\end{figure}
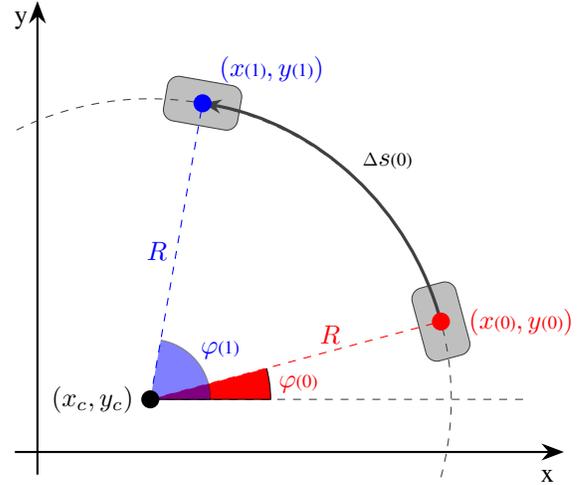

The length of the driven trajectory $\sm{$\Delta$} s\sm{(0)}$ between $(x\sm{(0)},y\sm{(0)})$ and $(x\sm{(1)},y\sm{(1)})$ is the circular arc and is represented by 
\begin{IEEEeqnarray}{lCl}
	\sm{$\Delta$} s\sm{(0)} &=& R \!\cdot\! (\varphi\sm{(1)} - \varphi\sm{(0)}) \label{eq:ds} 
\end{IEEEeqnarray}

Solve equation \eqref{eq:ds} for $\varphi\sm{(1)}$ and use the relation between radius $R$ and curvature $\kappa$ with $R=\tfrac{1}{\kappa}$. 
Thus $\varphi\sm{(1)}$ is given by
\begin{IEEEeqnarray}{lCl}
	\varphi\sm{(1)} &=& \varphi\sm{(0)} + \sm{$\Delta$} s\sm{(0)} \!\cdot\! \kappa \label{eq:phi1a} 
\end{IEEEeqnarray}

Expand equation \eqref{eq:x1b} and \eqref{eq:y1b}, using \eqref{eq:phi1a} is
\begin{IEEEeqnarray}{lCl}
	x\sm{(1)} &=& x\sm{(0)} - 2 \!\cdot\! R \!\cdot\! \sin\bigl(\tfrac{2 \cdot \varphi\sm{(0)} + \sm{$\Delta$} s\sm{(0)} \cdot \kappa}{2} \bigr) \!\cdot\! \sin\bigl(\tfrac{\sm{$\Delta$} s\sm{(0)} \cdot \kappa }{2} \bigr)
	\label{eq:x1c} \\
	y\sm{(1)} &=& y\sm{(0)} + 2 \!\cdot\! R \!\cdot\! \cos\bigl(\tfrac{2 \cdot \varphi\sm{(0)} + \sm{$\Delta$} s\sm{(0)} \cdot \kappa}{2} \bigr) \!\cdot\! \sin\bigl(\tfrac{\sm{$\Delta$} s\sm{(0)} \cdot \kappa }{2} \bigr) 
	\label{eq:y1c} 
\end{IEEEeqnarray}

$\sin\bigl(0.5 \!\cdot\! \sm{$\Delta$} s\sm{(0)} \!\cdot\! \kappa \bigr) \approx 0.5 \!\cdot\! \sm{$\Delta$} s\sm{(0)} \!\cdot\! \kappa$ can be assumed for small angles. This approximated equation \eqref{eq:x1c} and \eqref{eq:y1c} can written as
\begin{IEEEeqnarray}{lCl}
	x\sm{(1)} &\approx& x\sm{(0)} - \sm{$\Delta$} s\sm{(0)} \!\cdot\! \sin\bigl(\varphi\sm{(0)} + 0.5 \!\cdot\! \sm{$\Delta$} s\sm{(0)} \!\cdot\! \kappa \bigr) \label{eq:x1d} \\
	y\sm{(1)} &\approx& y\sm{(0)} + \sm{$\Delta$} s\sm{(0)} \!\cdot\! \cos\bigl(\varphi\sm{(0)} + 0.5 \!\cdot\! \sm{$\Delta$} s\sm{(0)} \!\cdot\! \kappa \bigr) \label{eq:y1d} 
\end{IEEEeqnarray}

Assuming that the vehicle drives between two time steps $\sm{$\Delta$} t = t_k - t_{k-1}$ with constant velocity $v$, distance $\sm{$\Delta$} s\sm{(0)}$ can written as 
\begin{IEEEeqnarray}{lCl}
	\sm{$\Delta$} s\sm{(0)} &=& v\sm{(0)} \!\cdot\! \sm{$\Delta$} t \label{eq:s0} 
\end{IEEEeqnarray}

Using this equation and substituting $\varphi\sm{(0)} + 0.5 \!\cdot\! \sm{$\Delta$} s\sm{(0)} \!\cdot\! \kappa$ with $\alpha\sm{(0)}$, equation \eqref{eq:x1d} and \eqref{eq:y1d} can be simplify as follows
\begin{IEEEeqnarray}{lCl}
	x\sm{(1)} &\approx& x\sm{(0)} - v\sm{(0)} \!\cdot\! \sm{$\Delta$} t \!\cdot\! \sin(\alpha\sm{(0)}) \label{eq:x1e} \\
	y\sm{(1)} &\approx& y\sm{(0)} + v\sm{(0)} \!\cdot\! \sm{$\Delta$} t \!\cdot\! \cos(\alpha\sm{(0)}) \label{eq:y1e} 
\end{IEEEeqnarray}

Using equation \eqref{eq:s0} and considering 
\begin{IEEEeqnarray}{lCl}
	\alpha\sm{(0)} &=& \varphi\sm{(0)} + 0.5 \!\cdot\! \sm{$\Delta$} s\sm{(0)} \!\cdot\! \kappa \nonumber \\
	&=& \varphi\sm{(0)} + 0.5 \!\cdot\! v\sm{(0)} \!\cdot\! \sm{$\Delta$} t \!\cdot\! \kappa \label{eq:alpha0} \\ 
	\alpha\sm{(1)} &=& \varphi\sm{(1)} + 0.5 \!\cdot\! \sm{$\Delta$} s\sm{(1)} \!\cdot\! \kappa \nonumber \\ 
	&=& \varphi\sm{(1)} + 0.5 \!\cdot\! v\sm{(1)} \!\cdot\! \sm{$\Delta$} t \!\cdot\! \kappa \label{eq:alpha1} 
\end{IEEEeqnarray}

equation \eqref{eq:phi1a} can written as
\begin{IEEEeqnarray}{rCl}
	\varphi\sm{(1)} &=& \varphi\sm{(0)} + v\sm{(0)} \!\cdot\! \sm{$\Delta$} t \!\cdot\! \kappa \nonumber \\
	\alpha\sm{(1)} - 0.5 \!\cdot\! v\sm{(1)} \!\cdot\! \sm{$\Delta$} t \!\cdot\! \kappa &=& \alpha\sm{(0)} - 0.5 \!\cdot\! v\sm{(0)} \!\cdot\! \sm{$\Delta$} t \!\cdot\! \kappa + v\sm{(0)} \!\cdot\! \sm{$\Delta$} t \!\cdot\! \kappa 
	\nonumber \\
	\alpha\sm{(1)} &=& \alpha\sm{(0)} + 0.5 \!\cdot\! \sm{$\Delta$} t \!\cdot\! \kappa \!\cdot\! (v\sm{(1)} + v\sm{(0)}) \label{eq:alpha1b} 
\end{IEEEeqnarray}

Equation \eqref{eq:alpha1b} can be simplify if the velocity between two time steps is approximately constant, thus $v\sm{(1)} \approx v\sm{(0)}$
\begin{IEEEeqnarray}{lCl}
	\alpha\sm{(1)} &\approx& \alpha\sm{(0)} + v\sm{(0)} \!\cdot\! \sm{$\Delta$} t \!\cdot\! \kappa \label{eq:alpha1c} 
\end{IEEEeqnarray}

Summing up, the following equation describes the model of moving vehicles (assuming small angles thus $\sin\bigl(0.5 \!\cdot\! \sm{$\Delta$} s\sm{(0)} \!\cdot\! \kappa \bigr) \approx 0.5 \!\cdot\! \sm{$\Delta$} s\sm{(0)} \!\cdot\! \kappa$): 
\begin{IEEEeqnarray}{lCl}
	x\sm{(i+1)}      &\approx& x\sm{(i)} - v\sm{(i)} \!\cdot\! \sm{$\Delta$} t\sm{(i)} \!\cdot\! \sin(\alpha\sm{(i)}) \label{eq:x1m} \\
	y\sm{(i+1)}      &\approx& y\sm{(i)} + v\sm{(i)} \!\cdot\! \sm{$\Delta$} t\sm{(i)} \!\cdot\! \cos(\alpha\sm{(i)}) \label{eq:y1m} \\
	\alpha\sm{(i+1)} &\approx& \alpha\sm{(i)} + v\sm{(i)} \!\cdot\! \sm{$\Delta$} t\sm{(i)} \!\cdot\! \kappa\sm{(i)} \label{eq:alpha1m} \\
	\kappa\sm{(i+1)} &\approx& \kappa\sm{(i)} \label{eq:kappa1m} \\
	v\sm{(i+1)}      &\approx& v\sm{(i)} \label{eq:v1m} 
\end{IEEEeqnarray}

Note that for many applications, term $0.5 \!\cdot\! \sm{$\Delta$} s\sm{(0)} \!\cdot\! \kappa$ is very small. In these cases angle is approximatly $\alpha \approx \varphi$.

\subsection{Linear model}
\label{ref:LinearModel}

Another way to describe kinematics of a vehicle can be achieved by using models which use setting a certain derivative of the position to zero \cite{BarShalom.2001} as basis.
For example, assuming that a vehicle moves between two time steps $\sm{$\Delta$} t$ with constant velocity $v$, the movement is given by
\begin{IEEEeqnarray}{lCl}
	x\sm{(i+1)}  &=& x\sm{(i)} + v_x\sm{(i)} \!\cdot\! \sm{$\Delta$} t\sm{(i)} \label{eq:xla} \\
	v_x\sm{(i+1)} &\approx& v_x\sm{(i)} \label{eq:vxla} \\
	y\sm{(i+1)}  &=& y\sm{(i)} + v_y\sm{(i)} \!\cdot\! \sm{$\Delta$} t\sm{(i)} \label{eq:yla} \\
	v_y\sm{(i+1)} &\approx& v_y\sm{(i)} \label{eq:vyla} 
\end{IEEEeqnarray}

where $v_x\sm{(i)}$ and $v_y\sm{(i)}$ are the velocity of x- and y-coordinate.

\section{Kalman Filter}
\subsection{Kalman Equations}
In general a physical system can be written as state-space model
\begin{IEEEeqnarray}{rCl}
	\underline{x}\sm{(i+1)} &=& \underline{f}\bigl(\underline{x}\sm{(i)},\underline{u}\sm{(i)}\bigr) + \underline{z}\sm{(i)}  \\
	\underline{y}\sm{(i)}   &=& \underline{h}\bigl(\underline{x}\sm{(i)},\underline{u}\sm{(i)}\bigr) + \underline{\omega}\sm{(i)} 
\end{IEEEeqnarray}

where $\underline{f}\bigl(\underline{x}\sm{(i)},\underline{u}\sm{(i)}\bigr)$ and $\underline{h}\bigl(\underline{x}\sm{(i)},\underline{u}\sm{(i)}\bigr)$ are two (nonlinear) equation vectors. 
$\underline{x}\sm{(i)}$ is the state vector, $\underline{u}\sm{(i)}$ is the control vector and $\underline{y}\sm{(i)}$ is the measurements (or observations) at time step k. 
Model uncertainties are illustrated with the random vector $\underline{z}\sm{(i)}$ and errors of measurements are described by the random vector $\underline{\omega}\sm{(i)}$.

A Kalman Filter for such a state-space representation is given by the following equations
\begin{IEEEeqnarray}{rCl}
	\underline{K}\sm{(i)} &=& \underline{\hat{P}}\sm{(i)} \!\cdot\! \underline{C}_{\,J}^T \!\cdot\! \bigl( \underline{C}_{\,J} \!\cdot\! \underline{\hat{P}}\sm{(i)} \!\cdot\!
	\underline{C}^T_{\,J} + \underline{R}\sm{(i)} \bigr)^{-1} \\
	\underline{\tilde{x}}\sm{(i)} &=& \underline{\hat{x}}\sm{(i)} + \underline{K}\sm{(i)} \!\cdot\! \Bigl(\underline{y}\sm{(i)} - \underline{h}\bigl(\underline{\hat{x}}\sm{(i)},\underline{u}\sm{(i)}\bigr)\Bigr) \\
	\underline{\tilde{P}}\sm{(i)} &=& \bigl(\underline{I}-\underline{K}\sm{(i)} \!\cdot\! \underline{C}_{\,J}\bigr) \!\cdot\! \underline{\hat{P}}\sm{(i)} \!\cdot\! \bigl(\underline{I}-\underline{K}\sm{(i)} \!\cdot\! \underline{C}_{\,J}\bigr)^T \nonumber \\
	&& + \underline{K}\sm{(i)} \!\cdot\! \underline{R}\sm{(i)} \!\cdot\! \underline{K}\sm{(i)}^T \vspace{1mm} \\
	\underline{\hat{x}}\sm{(i+1)} &=& \underline{f}\bigl(\underline{\tilde{x}}\sm{(i)},\underline{u}\sm{(i)}\bigr)  \\ 
	\underline{\hat{P}}\sm{(i+1)} &=& \underline{A}_{\,J} \!\cdot\! \bigl(\underline{\tilde{P}}\sm{(i)} + \underline{Q}\sm{(i)} \bigr) \!\cdot\! \underline{A}^T_{\,J}
\end{IEEEeqnarray}

with $\underline{Q}\sm{(i)}=\!\text{Var}\bigl( \underline{z}\sm{(i)} \bigr)$, $\underline{R}\sm{(i)}=\!\text{Var}\bigl(\underline{\omega}\sm{(i)}\bigr)$ and the Jacobian matrices $\underline{A}_{\,J}$ and $\underline{C}_{\,J}$ 
\begin{IEEEeqnarray}{C}
	\underline{A}_{\,J} \!=\! \begin{bmatrix} \dfrac{\partial\underline{f}}{\partial\tilde{x}_1} & \!\!\cdots\!\! & \dfrac{\partial\underline{f}}{\partial\tilde{x}_n} \end{bmatrix},
	\hspace{1mm} 
	\underline{C}_{\,J} \!=\! \begin{bmatrix} \dfrac{\partial\underline{h}}{\partial\tilde{x}_1} & \!\!\cdots\!\! & \dfrac{\partial\underline{h}}{\partial\tilde{x}_n} \end{bmatrix}
\end{IEEEeqnarray}

\subsection{Nonlinear model - Extended Kalman Filter}

Choosing the state vector $\underline{x}\sm{(i)}$ as
\begin{IEEEeqnarray}{C}
	\underline{x}\sm{(i)} = \begin{bmatrix} x\sm{(i)} & y\sm{(i)} & \alpha\sm{(i)} & \kappa\sm{(i)} & v\sm{(i)} \end{bmatrix}^T
\end{IEEEeqnarray}

and using the nonlinear model described in section \ref{ref:NonlinearModel} the Jacobian matrices $\underline{A}_{\,J}$ and $\underline{C}_{\,J}$ is written as
\begin{IEEEeqnarray}{rCl}
	\underline{A}_{J} \!&=&\!\! 
	\begin{bmatrix} 
		1 & \ha 0 \ha & \mm v\sm{(i)} \!\cdot\! \sm{$\Delta$} t\sm{(i)} \!\cdot\! \cos\bigl(\alpha\sm{(i)}\bigr) & 0 & \!\mm\sm{$\Delta$} t\sm{(i)} \!\cdot\! \sin\bigl(\alpha\sm{(i)}\bigr) \\
		0 & \ha 1 \ha & \mm v\sm{(i)} \!\cdot\! \sm{$\Delta$} t\sm{(i)} \!\cdot\! \sin\bigl(\alpha\sm{(i)}\bigr) & 0 & \!\sm{$\Delta$} t\sm{(i)} \!\cdot\! \cos\bigl(\alpha\sm{(i)}\bigr) \\
		0 & \ha 0 \ha & 1 & \hd v\sm{(i)} \!\cdot\! \sm{$\Delta$} t\sm{(i)} \hd & \sm{$\Delta$} t\sm{(i)} \!\cdot\! \kappa\sm{(i)}  \\
		0 & \ha 0 \ha & 0 & 1 & 0 \\
		0 & \ha 0 \ha & 0 & 0 & 1 
	\end{bmatrix} \vspace{1mm}\\
	\underline{C}_{J} \!&=&\! \begin{bmatrix} 1 & 0 & 0 & 0 & 0 \\ 0 & 1 & 0 & 0 & 0 \end{bmatrix}
\end{IEEEeqnarray}

and
\begin{IEEEeqnarray}{rCl}
	\underline{z}\sm{(i)} &=& \begin{bmatrix} z_1\sm{(i)} & z_2\sm{(i)} & z_3\sm{(i)} & z_4\sm{(i)} & z_5\sm{(i)} \end{bmatrix}^T
\end{IEEEeqnarray}

\subsection{Covariance of Measurement Noise}
\label{ref:adaptiveR}

The Covariance of measurement noise can be calculated by
\begin{IEEEeqnarray}{rCl}
	\underline{R}\sm{(i)} &=& \text{Var}\bigl(\underline{\omega}\sm{(i)}\bigr) = \text{Var}\bigl(\underline{y}\sm{(i)}\bigr) \\
	\underline{R}\sm{(i)} &=& \text{E}\Bigl(\bigl(\underline{y}\sm{(i)} - \text{E}\bigl(\underline{y}\sm{(i)}\bigr)\bigr) \!\cdot\! \bigl(\underline{y}\sm{(i)} - \text{E}\bigl(\underline{y}\sm{(i)}\bigr)\bigr)^T \Bigr) \label{eq:R1} 
\end{IEEEeqnarray}

The linear model described in section \ref{ref:LinearModel} can be used to estimate the expected value $\text{E}\bigl(\underline{y}\sm{(i)}\bigr)$. In this linear model each coordinate is independent of the other one, so it is better to use one Kalman Filter for each coordinate.    

Choosing the state vector $\underline{x}\sm{(i)}$ as
\begin{IEEEeqnarray}{C}
	\underline{x}\sm{(i)} = \begin{bmatrix} x\sm{(i)} & v_x\sm{(i)} \end{bmatrix} \quad \text{or} \quad 
	\underline{x}\sm{(i)} = \begin{bmatrix} y\sm{(i)} & v_y\sm{(i)} \end{bmatrix}
\end{IEEEeqnarray}

the matrices $\underline{A}_{\,J}$ and $\underline{C}_{\,J}$ is written as
\begin{IEEEeqnarray}{C}
	\underline{A}_{J} = \begin{bmatrix} 1 & \sm{$\Delta$} t\sm{(i)} \\ 0 & 1 \end{bmatrix}^T \quad \underline{C}_{J} = \begin{bmatrix} 1 & 0 \end{bmatrix}
\end{IEEEeqnarray}

Assuming a time invariant variance of random vector $\underline{z}\sm{(i)}$ and $\underline{\omega}\sm{(i)}$ the update of the state vector for this linear model can be simplified as (see \cite{Marchthaler.2017}).

\begin{IEEEeqnarray}{C}
	\text{E}\bigl(\underline{y}\sm{(i)}\bigr) = \underline{\tilde{x}}(k) = \underline{y}(k) + \bigl( \underline{I} - \underline{K} \!\cdot\! \underline{C}_J\bigr) \!\cdot\! \underline{A}_J \!\cdot\! \underline{\tilde{x}}(k-1)
\end{IEEEeqnarray}

with 
\begin{IEEEeqnarray}{C}
	K \!=\! \frac{0.125}{\sm{$\Delta$} t\sm{(i)}} \!\cdot\! \begin{bmatrix} 
		\sm{$\Delta$} t\sm{(i)} \!\cdot\!(\sn{$-\lambda^2 - 8 \!\cdot\!\! \lambda + (\lambda + 4) \!\cdot\! \sqrt{\lambda^2 + 8 \!\cdot\!\! \lambda)}$} \;\\
		\sn{$2 \!\cdot\! (\lambda^2 + 4 \!\cdot\!\! \lambda - \lambda \!\cdot\! \sqrt{\lambda^2 + 8 \!\cdot\!\! \lambda}$})
	\end{bmatrix} \\
	\lambda = \sm{$\Delta$} t\sm{(i)} \cdot \sqrt{\tfrac{Q}{R}} 
\end{IEEEeqnarray}

Assuming that the covariance of measurement noise will change slowly, equation \eqref{eq:R1} can be written as (see \cite{Marchthaler.2017})

\begin{IEEEeqnarray}{C}
	\underline{R}\sm{(i)} = \gamma \cdot \alpha_R \cdot \bigl(\text{E}\bigl(\underline{y}\sm{(i)}\bigr) - \underline{y}\sm{(i)}\bigr) \cdot \bigl(\text{E}\bigl(\underline{y}\sm{(i)}\bigr) - \underline{y}\sm{(i)}\bigr)^T \nonumber \\ + (1-\alpha_R) \cdot \underline{R}\sm{(i-1)}
\end{IEEEeqnarray}

\vspace*{3mm} 
\section{Results}
\subsection{Position}

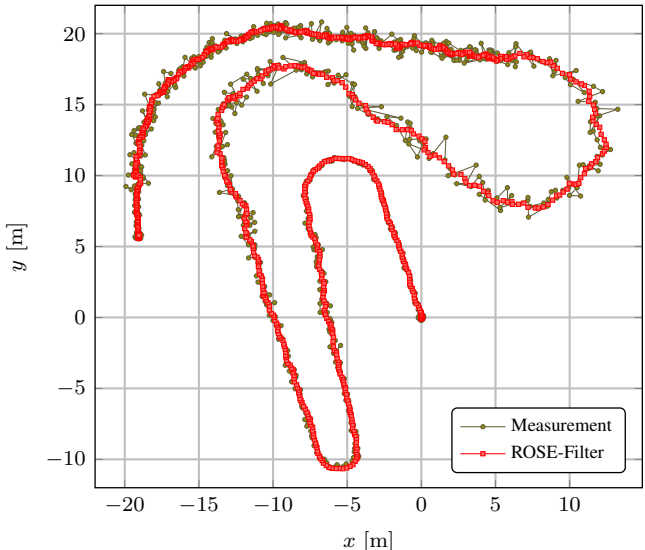
\begin{figure}[ht]
	\begin{center}
		\begin{tikzpicture}
			\begin{axis}[width=\columnwidth, height=80mm, xmin={-22}, xmax={14.9}, ymin={-12}, ymax={22}, x unit=\unit{m}, y unit=\unit{m},
				xlabel={$x$}, ylabel={$y$},	xtick = {-25,-20,...,15}, ytick = {-15,-10,...,25}, legend pos=south east, legend style={font=\scriptsize}] 
				\addplot[olive!50!black, thin, mark=*, mark options={fill=olive!95!black, solid, scale=0.4}] table[x={Pos_x}, y={Pos_y}] {EKF_ROSE.dat};
				\addplot[red, thin, mark=square*, mark options={fill=red!60, solid, scale=0.4}] table[x={ROSE_x}, y={ROSE_y}] {EKF_ROSE.dat};
				\legend{Measurement,ROSE-Filter}
			\end{axis}
		\end{tikzpicture}
		\vspace*{-3mm}
		\caption{Input data and filtered position with ROSE-Filter}
		\label{fig_inputdata}
	\end{center}
\end{figure}

To improve the performance of the ROSE-Filter a vehicle with changing velocity is used.
The covariance of measurement noise change over the time. 
(See olive line in fig. \ref{fig_inputdata}). The red line shows the ROSE-filtered x- and y-position of the vehicle. 

\vspace*{2mm}

\newpage
\subsection{Orientation, Curvature and Velocity}

\begin{figure}[ht]
	\begin{flushright}
		\begin{tikzpicture}
			\begin{axis}[width=.98\columnwidth, height=55mm, xmin=0, xmax=100, xlabel={time}, ylabel={orientationn ($\alpha$)}, x unit=\unit{s}, y unit=\unit{rad},
				legend pos=south east, legend style={font=\scriptsize}, 
				xticklabel style={/pgf/number format/precision=0, /pgf/number format/fixed, /pgf/number format/fixed zerofill},
				yticklabel style={/pgf/number format/precision=1, /pgf/number format/fixed, /pgf/number format/fixed zerofill}]
				\addplot[blue] table[x={time}, y={EKF_alpha}]{EKF_ROSE.dat};
				\addplot[red] table[x={time}, y={ROSE_alpha}]{EKF_ROSE.dat};
				\legend{Classical EKF,ROSE-Filter}
			\end{axis} 
		\end{tikzpicture}
		\begin{tikzpicture}
			\begin{axis}[width=.98\columnwidth, height=55mm, xmin=0, xmax=100, xlabel={time}, ylabel={curvature ($\kappa$)}, x unit=\unit{s}, y unit=\unitfrac{1}{m}, 
				legend pos=south east, legend style={font=\scriptsize},
				xticklabel style={/pgf/number format/precision=0, /pgf/number format/fixed, /pgf/number format/fixed zerofill},
				yticklabel style={/pgf/number format/precision=1, /pgf/number format/fixed, /pgf/number format/fixed zerofill}]
				\addplot[blue] table[x={time}, y={EKF_Kr}]{EKF_ROSE.dat};
				\addplot[red] table[x={time}, y={ROSE_Kr}]{EKF_ROSE.dat};
				\legend{Classical EKF,ROSE-Filter}
			\end{axis} 
		\end{tikzpicture}
		\begin{tikzpicture}
			\begin{axis}[width=.98\columnwidth, height=55mm, xmin=0, xmax=100, xlabel={time}, ylabel={velocity ($v$)}, x unit=\unit{s}, y unit=\unitfrac{m}{s},
				legend pos=north west, legend style={font=\scriptsize}, ytick = {0,1,...,4},
				xticklabel style={/pgf/number format/precision=0, /pgf/number format/fixed, /pgf/number format/fixed zerofill}, 
				yticklabel style={/pgf/number format/precision=1, /pgf/number format/fixed, /pgf/number format/fixed zerofill}]
				\addplot[blue] table[x={time}, y={EKF_v}]{EKF_ROSE.dat};
				\addplot[red] table[x={time}, y={ROSE_v}]{EKF_ROSE.dat};
				\legend{Classical EKF,ROSE-Filter}
			\end{axis}
		\end{tikzpicture}
		\caption{Estimated orientation, curvature and velocity of vehicle with classical EKF and ROSE-Filter}
		\label{fig_result}
	\end{flushright}
\end{figure}
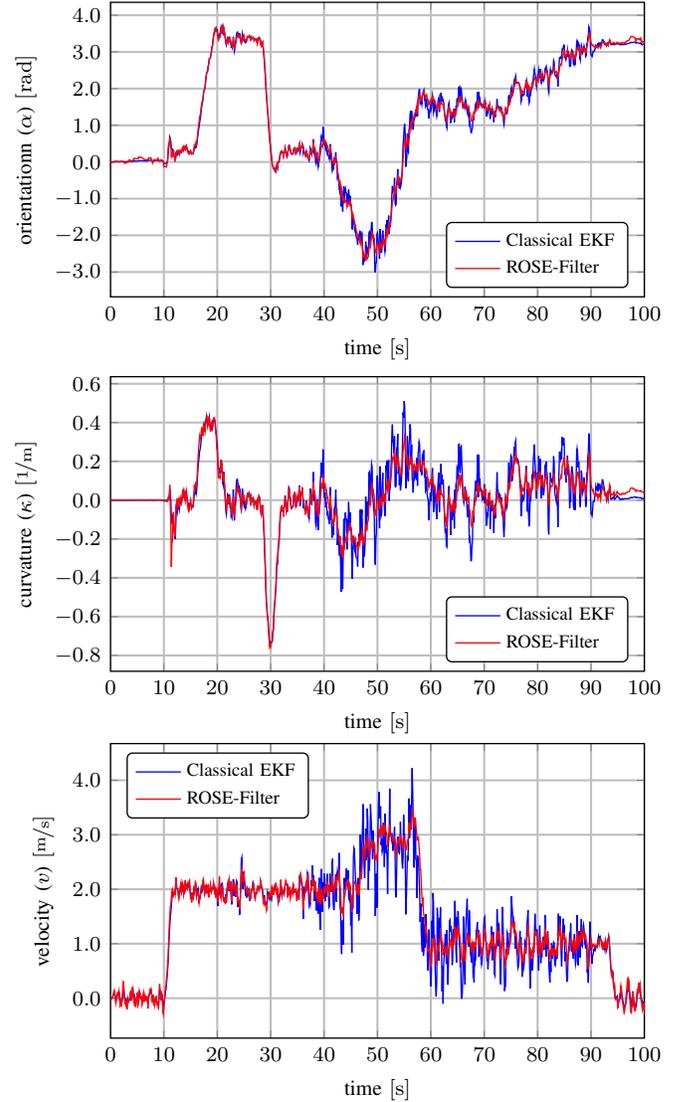

Fig. \ref{fig_result} compares the results of two different approaches regarding the estimated values, orientation of the vehicle, curvature of the driven course and the velocity.
The classical EKF is based on the model described in section \ref{ref:LinearModel} with a static covariance of measurement noise which is determined at the beginning of the driving cycle.
Whereas the ROSE-Filter is based on the same physical model but with an adaptive covariance of measurement noise described in section \ref{ref:adaptiveR}. 

\vspace*{2mm}

The estimated values calculated with ROSE-Filter are smoother and growth more weakly with increasing noise of the measured position.

\newpage
\subsection{Root Mean Square}

To determine the quality of the classical EKF and the ROSE-Filter the root-mean-square error is calculated for each estimated value.

\begin{IEEEeqnarray}{C}
	\text{RMS Errors} = \sqrt{\frac{\sum\limits_{i=1}^{N}{(\tilde{x}\sm{(i)}-{x}\sm{(i)})^2}}{N}} 
\end{IEEEeqnarray}

The root-mean-square error of the position is calculated by the euclidean distance. 

In tab. \ref{tab_rms} the root-mean-square error for each estimated value is shown for the test drive of the vehicle. 

\vspace*{2mm}
\begin{table}[ht]
	\begin{center}
		\begin{tabular}{lcccc}
			\toprule
			&\multicolumn{4}{c}{$\text{RMS Errors}$} \\[1mm]
			& position  & orientation  & curvature  & velocity \\ 
			& ($x$,$y$) &  ($\alpha$)  & ($\kappa$) &  ($v$)   \\ 
			\midrule
			EKF         & 0.215 & 0.236 & 0.142 & 0.332 \\
			ROSE-Filter & 0.182 & 0.183 & 0.123 & 0.223 \\ 
			\midrule
			Improvement & 18.2\%  & 28.6\%  & 15.4\%  & 48.5\% \\
			\bottomrule
		\end{tabular}
		\caption{root-mean-square error and improvement of ROSE-Filter compared to classical EKF}
		\label{tab_rms}
	\end{center}
\end{table}

The improvement of the ROSE-Filter is between 15.4\% and 48.5\% depending on the estimated value. 
An averaged enhancement of 27.7\% is achieved.

\vspace*{3mm}
\section{Conclusion}
This paper illustrates the way for estimating position, orientation and velocity of a vehicle by measuring its x- and y-position.
For this, a Extended Kalman-Filter (EKF) uses a model which calculates the trajectory based on circular path. With such a model the given position can be smoothed and the orientation of the vehicle, curvature of the driven course and the velocity can be estimated.

Changing in the covariance of measurement noise can be determined continuously by using a ROSE-Filter. Thus for every measured x- and y-position the correct covariance of measurement noise is available . 






\bibliography{EKF_xy}

\begin{thebibliography}{1}
\providecommand{\url}[1]{#1}
\csname url@samestyle\endcsname
\providecommand{\newblock}{\relax}
\providecommand{\bibinfo}[2]{#2}
\providecommand{\BIBentrySTDinterwordspacing}{\spaceskip=0pt\relax}
\providecommand{\BIBentryALTinterwordstretchfactor}{4}
\providecommand{\BIBentryALTinterwordspacing}{\spaceskip=\fontdimen2\font plus
\BIBentryALTinterwordstretchfactor\fontdimen3\font minus
  \fontdimen4\font\relax}
\providecommand{\BIBforeignlanguage}[2]{{%
\expandafter\ifx\csname l@#1\endcsname\relax
\typeout{** WARNING: IEEEtran.bst: No hyphenation pattern has been}%
\typeout{** loaded for the language `#1'. Using the pattern for}%
\typeout{** the default language instead.}%
\else
\language=\csname l@#1\endcsname
\fi
#2}}
\providecommand{\BIBdecl}{\relax}
\BIBdecl

\bibitem{Lugner.2019}
P.~Lugner, \emph{\BIBforeignlanguage{english}{Vehicle Dynamics of Modern
  Passenger Cars}}.\hskip 1em plus 0.5em minus 0.4em\relax Springer, Berlin,
  Heidelberg: Springer, 2019.

\bibitem{Schramm.2014}
D.~Schramm, M.~Hiller, and R.~Bardini,
  \emph{\BIBforeignlanguage{english}{Vehicle Dynamics of Modern Passenger
  Cars}}.\hskip 1em plus 0.5em minus 0.4em\relax Berlin, Heidelberg: Springer,
  2014.

\bibitem{BarShalom.2001}
Y.~Bar-Shalom, T.~Kirubarajan, and X.-R. Li, \emph{Estimation with applications
  to tracking and navigation - theory algorithms and software}.\hskip 1em plus
  0.5em minus 0.4em\relax New York, Chichester, Weinheim, Brisbane, Singapore,
  Toronto: John Wiley \& Sons, Inc., 2001.

\bibitem{Marchthaler.2017}
R.~Marchthaler and S.~Dingler, \emph{Kalman-Filter - Einf{\"u}hrung in die
  Zustandssch{\"a}tzung und ihre Anwendung f{\"u}r eingebettete Systeme}.\hskip
  1em plus 0.5em minus 0.4em\relax Springer Vieweg, Wiesbaden, 2017.

\end{thebibliography}

\begin{IEEEbiography}[{\includegraphics[width=1in,height=1.25in,clip,keepaspectratio]{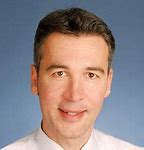}}]
{Reiner Marchthaler} received his graduate engineer of 
communications engineering from the University of Applied Science in Ulm in 1994 and his graduate
engineer of electrical engineering and the Ph.D. from the University of Hagen in 2001 and 2004, 
respectively. 

He joined Robert Bosch GmbH, Germany, in 1994. He was a system-engineer for adaptive cruise 
control system (Advanced Driver Assistance Systems) between 1994 and 1998. Afterwards he was 
project-engineer for an occupant classification systems for an adaptive restraint system. 
In 2004 he joined the research center of Robert Bosch and developed a research group for combined
active and passive safety, pedestrian protection and accident research. 

R. Marchthaler has a professorship since 2011 of embedded systems with the area of expertise
“autonomous systems” at University of Applied Sciences Esslingen

He published many technical memoranda, journal and conference papers, 
and is one of the two authors of the book "Kalman-Filter: Einführung in die Zustandsschätzung und ihre 
Anwendung für eingebettete Systeme", Springer Vieweg, Wiesbaden , 2017, ISBN 978–3658167271.
He also has more than 100 (pending) patents. 
\end{IEEEbiography}



\end{document}